\documentclass[12pt,a4paper]{article}
\usepackage{graphicx}
\usepackage{color}
\usepackage{cite}
\usepackage{soul}
\usepackage{amssymb}
\usepackage[normalem]{ulem}
\usepackage{cancel}
\usepackage{parallel}
\usepackage{color}
\usepackage[reqno]{amsmath}
\usepackage{array}
\usepackage{tabulary}
\usepackage{wasysym}
\usepackage{rotating}
\usepackage[T1]{fontenc} 
\usepackage[utf8]{inputenc}
\usepackage{yfonts} 
\usepackage[cc]{titlepic}
\usepackage{todonotes}
\usepackage{hyperref}  

\def\be{\begin{equation}}
\def\ee{\end{equation}}
\def\bea{\begin{eqnarray}}
\def\eea{\end{eqnarray}}

\begin{document}

\begin{titlepage}
\title{Ising's roots and the  transfer-matrix eigenvalues}
\author{\small Reinhard Folk\thanks{e-mail: r.folk@liwest.at} \\ \small Institute for Theoretical Physics, Johannes Kepler University Linz, \\
\small 4040 Linz, Austria \and 
\small Yurij Holovatch\thanks{e-mail: hol@icmp.lviv.ua} \\ \small Institute for Condensed Matter Physics, 
National Acad. Sci. of Ukraine, \\ \small 79011 Lviv, Ukraine\\
\small 
$\mathbb{L}^4$ Collaboration and Doctoral College for the Statistical Physics of  \\
\small Complex Systems, Lviv-Leipzig-Lorraine-Coventry, Europe\\
\small Centre for Fluid and Complex Systems, Coventry University, \\
\small 			Coventry CV1 5FB, UK \\
\small Complexity Science Hub Vienna, 1080 Vienna, Austria
} 
\end{titlepage}
\maketitle
\begin{abstract}
Today, the Ising model is an archetype describing collective ordering processes. 
And, as such, it is widely known in physics and far beyond. Less known is the fact 
that the thesis defended by Ernst Ising 100 years ago (in 1924) contained not only 
the solution of what we call now  the `classical 1D Ising model' but also other problems. Some of 
these problems, as well as the method of their solution, are the subject of this 
 note. In particular, we discuss the combinatorial method Ernst Ising used 
to calculate the partition function for a chain of elementary magnets. In the thermodynamic limit,
this method leads to the result that the partition function is given by the roots of a 
certain polynomial. We explicitly show that  `Ising's roots'
 that arise within the combinatorial treatment  are also recovered  by the eigenvalues of the transfer matrix,
a concept that was introduced much later.
Moreover, we discuss the generalization of the two-state model to a three-state one
presented in Ising's  thesis, but not included in his famous paper of 1925
({\it E. Ising, Z. Physik {\bf 31} (1925) 253}).
The latter model can be considered as a forerunner of the now abundant models with many-component order parameters.
\end{abstract}

Keywords: Ising model; Ernst Ising; transfer matrix; Potts model

\section{Introduction}
The now famous Ising model has been suggested  by Wilhem Lenz to his student Ernst Ising in 1922. It was solved
	in 1D and presented in 1924 in Ernst Ising doctoral thesis \cite{Ising24} that was followed by a paper in 1925 \cite{Ising25}.
	The model has been considered by Ising without referring to the Hamiltonian. The latter, in the form as we know it today, 
	has been written down by Wolfgang  Pauli in  1930 \cite{Pauli30}. Instead, Ising defined weights of different configurations 
	and used combinatorial approach to calculate their contributions to the partition function. 
	In order to proceed, he introduced an auxiliary function leading to a polynomial whose roots allowed to 
	calculate the partition function. In the thermodynamic limit one only needs to know the largest root 
	to calculate all thermodynamic functions \cite{Bitter37}.

	In our days, it is a textbook exercise to calculate the partition function of a chain of two-state
elements with nearest neighbour interaction, what we call a 1D Ising model now. Usually it is done
by applying the transfer matrix to calculate the sum of exponential functions with the Ising model Hamiltonian.
The transfer-matrix method was introduced in 1941 by Kramers, Wannier, and Montroll  \cite{Kramers41,Montroll41}. The eigenvalues of the transfer matrix allow to 
	obtain the partition function.  The difference in the combinatorial (used by Ising) and the transfer-matrix methods to calculate the partition 
	function lies in the treatment of the Boltzmann weights. Ising concentrates on the Boltzmann weight of a 
	configuration of the whole chain, whereas the transfer-matrix method concentrates on the weights of 
	interacting units.

The goal of our paper is to attract attention to the important (maybe surprising) point that 
	the polynomial found in treatment of configurations within the combinatorial approach appears to 
	be the one that follows from the secular equation within the transfer-matrix approach that treats 
	elementary units rather than looking on their configurations. We explicitly show that the `Ising's roots'
	that arise within the combinatorial treatment  are given by the eigenvalues of the transfer matrix. 
	Moreover, we will consider in more detail the Ising's  solution for the linear chain with magnetic elements
	allowing, besides  two, parallel and antiparallel, also transverse positions. Although this solution was displayed 
	in Ising's thesis \cite{Ising24} (see Figure 1),  it was not presented in his paper \cite{Ising25}. This is one of the reasons that 
	it is less familiar. Doing so, we will show that
	Ising's thesis not only contained the definition of what is called today the Ising model but 
	also the three-state model which can be considered as a  forerunner to the models with many-component
 order parameter, the Potts model being one of them. 
	This extension of the two-state model
	was solved by Ising only making assumptions which allowed him to present analytic results. Independent of these approximations he calculated an exact equation: a polynomial, whose largest root gives the partition function in the thermodynamic limit.

\begin{figure}\centering
\includegraphics[height=0.7\textwidth]{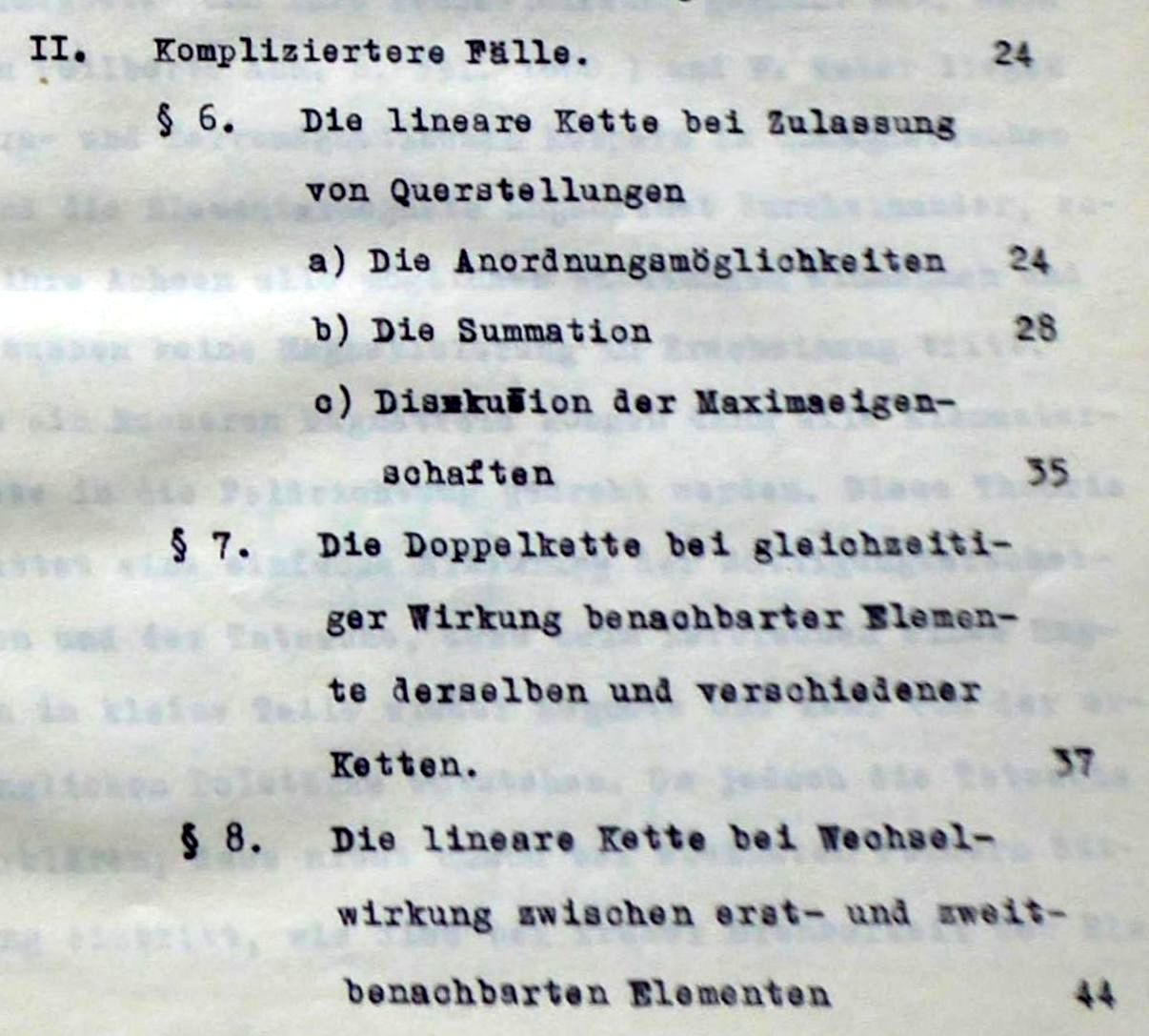}
\caption{A page from the table of content of Ernst Ising's doctoral thesis \cite{Ising24} featuring
`complicated cases', Komplizierte F\"alle ({\it Germ.}). One of these cases -- The linear chain when transverse positions are permitted (Die lineare Kette bei Zulassung von Querstellungen ({\it Germ.})) -- we discuss in this paper.
\label{fig1}}
\end{figure}

The set-up of the rest of the paper is the following: in the next section \ref{II}
	we will sketch Ising's solution of the  one-dimesional  two-state model. Then, in section 
	\ref{III} we will consider Ising's  solution for the linear chain with magnetic elements
	allowing also transverse positions. In turn, in section \ref{IV} we will show that Ising's three-state chain (the  case when only a single transfer position 
	is allowed)  relates to what is currently known as $q=3$-state Potts model.
	The last has been introduced much later in 1952 \cite{Potts52} and as far as we know the fact that 
	its forerunner  has been solved for $d=1$ as early as in 
	1924 has not been acknowledged \cite{Wu82}. We end by conclusions in section \ref{V}.

\section{Ising's method and the solution for a two-state model}\label{II} 

As noted in the Introduction,  neither the Hamiltonian nor the transfer matrix
	were used  in the original work of Ising, they were not even known(!), 
 see e.g. \cite{Brush67,Kobe00,Niss05,Folk22} for a more detailed history.  In this chapter we 
	briefly explain the method used in the original publication and show how it relates to the 
	now standard transfer matrix technique.

\subsection{Definition of states, configurations, and energy places}\label{II.1} 
\begin{figure}[h]
\centering \includegraphics[height=3.0cm]{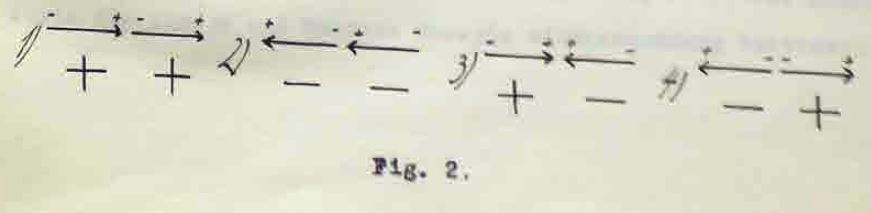}
\caption{\label{fig2} Possible mutual orientations of two neighbouring elementary magnets, original figure
from Ernst Ising's thesis \cite{Ising24}. Contributions to energy come from the places
where neighboring magnetic moments have opposite orientations,
panels {\em{3,4}}. Such places are called  {\em{energy places}} - {\em{Energiestelle (Germ.)}}.
}
\end{figure}

In his thesis, Ernst Ising follows an approach of statistical mechanics, developed by the time by 
Josiah Willard Gibbs and Ludwig Boltzmann. First, he considers a chain of $N$ elementary magnets  
in an external magnetic field when each of the magnets can be in two states, 
left/right or plus/minus as shown in Fig. \ref{fig2}. 
Central quantity of interest that defines thermodynamics of 
such system in equilibrium is its partition function, defined by:
\be \label{1}
Z=\sum_{\{ \rm configurations \} } e^{- \frac{\cal E}{kT}}
\ee
where $T$ is temperature, $k$ is Boltzmann constant. The sum spans all possible combinations of states of the magnetic elements, 
	i.e. all configurations of the elementary magnets in the chain, and energy ${\cal E}$ depends on the configuration. 
It should be noted the energy may have the same value for different configurations. 

 Ising argues that, for a given configuration of elementary magnets along a chain, 
contributions to its 
energy are of different origin: (i) due to alignment of the  magnetic moments along or opposite
to the field direction and (ii) due to the mutual orientation of the neighbouring elementary
magnets. Assuming that the same orientations of neighboring magnetic moments (cf. panels {\em{1,2}} 
in Fig. \ref{fig2}) take minimal energy (chosen to be zero), Ising arrives at the conclusion that only the
places where oppositely oriented magnets meet contribute to the energy. He calls such places the 
{\em{energy places}} (Energiestelle) (cf. panels {\em{3,4}} in Fig. \ref{fig2}). Consequently, the total
energy of a given configuration of $N$ elementary magnets is governed by three quantities:
 number of magnetic moments oriented along (or opposite to) the field, further denoted as
 $\nu_1$ ($\nu_2$), and number of the energy places $\sigma$. 
 The number ${\cal N} (\nu_1, \nu_2, \sigma)$ of different 
configurations that share the same values of $\nu_1$, $\nu_2$, $\sigma$ defines degeneracy of a microstate:
all such configurations have the same energy. The expression for the partition function (\ref{1}) of a chain
of length $N$ readily follows:
\be \label{2}
Z({N})=\sum_{\nu_1+\nu_2=N}\sum_{\sigma=0}^{N-1}{\cal N}(\nu_1,\nu_2,\sigma)e^{-\frac{E_\alpha(\nu_1,\nu_2)+E_\beta(\sigma)}{kT}} \, ,
\ee
with $E_\alpha(\nu_1,\nu_2)/(kT)=\alpha(\nu_2-\nu_1), \, \alpha= \mu h/(kT)$, $E_\beta(\sigma)/(kT)= \sigma \beta, \, \beta=e/(kT)$, $\mu$, $h$ are elementary magnetic moment
and an external magnetic field and $e$ is the energy value of a single energy place.
In the notations, Ernst Ising used in his thesis, this can be rewritten by introducing the Boltzmann weights
\be
A_1=e^\alpha, \, A_2=e^{-\alpha}, \, B=e^{-\beta} \,. \label{2a}
\ee
Here, $A_1, \, A_2$ are the weights for the states where the elementary magnets are  parallel or antiparallel  to the external field, and $B$ is the weight for the energy place of neighbouring antiparallel elementary magnet.  The energy scale has been chosen by Ising in such a way that  the Boltzmann weight of an energy place for parallel elementary magnets  is equal to one.
This leads to the following expression for the partition function (\ref{2}):
\be  \label{2b}
Z({N})=\sum_{\nu_1+\nu_2=N}\sum_{\sigma=0}^{N-1}{\cal N}(\nu_1,\nu_2,\sigma)A_1^{\nu_1}A_2^{\nu_2}B^\sigma\, .
\ee

\subsection{Ising's solution of the two-state model. An auxiliary function}\label{II.2} 
Ising used methods of combinatorics to explicitly count the number ${\cal N}(\nu_1,\nu_2,\sigma)$ 
expressing it via the binomial coefficients:\footnote{Ising's method to calculate $\cal{N}$ was mentioned in Lenz' review of Ising's thesis \cite{Folk22} and later cited in 1942 by T. S. Chang, Ph.D. and C. C. Ho, B.Sc., two former students of R. H. Fowler \cite{chang42}}
\be \label{3}
{\cal N}(\nu_1,\nu_2,\sigma=2s+\delta)= \binom{\nu_1-1}{s}\binom{\nu_2-1}{s+\delta-1} +\binom{\nu_2-1}{s}\binom{\nu_1-1}{s+\delta-1} \, ,
\ee
where $\delta$ takes on values $0$ or $1$ depending on the states of the chain boundary elements.
Furthermore, in order to proceed he suggested  to get rid of the condition 
	$\nu_1+\nu_2=N$ in (\ref{2}) since he was interested in the thermodynamic limit of the partition function. 
 To this end, he introduced an auxiliary function
\be \label{4}
F({x})=\sum_{{N}=0}^\infty Z({N}){x}^{N}\, .
\ee
Given an explicit form for $Z(N)$, Eqs. (\ref{2}), (\ref{3}), one can perform
the summation in (\ref{4}) arriving at:
\be  \label{5a}
{F}({x})=\frac{x[A_1+A_2-2A_1A_2(1-B)x]}{1-(A_1+A_2)x+A_1A_2(1-B^2)x^2} \, ,
\ee
and simplified as
\be\label{5b}
{F}({x})=\frac{2{x}[\cosh\alpha-(1-e^{-\beta}){x}]}{1-2\cosh\alpha\cdot {x}+(1-e^{-2\beta}){x}^2} \, .
\ee
Noteworthy, re-expanding $F(x)$ in terms of $x$ one recovers as expansion coefficients the partition function for finite $N$ at free boundary conditions. In the following the important quantities are the  inverse roots $w_1$ and $w_2$ of the polynomial in $x$ in the denominator of Eq. (\ref{5b}).
Indeed, this function can be put into series with respect to $x$:
\be \label{5c}
{F}({x})= \sum_{l=0}^\infty\left(a_1 w_1^{l}+ a_2 w_2^{l}\right){x}^l\, ,
\ee
with known explicit expressions for $a_i$ and $w_i$. In particular
\be \label{6}
w_{1\, ,2}=\cosh \alpha\pm \sqrt{\sinh^2\alpha +e^{-2\beta}}\, .
\ee
Observing that $w_1>w_2$ and comparing Eqs. (\ref{5c}) and (\ref{4}) one 
concludes that the exact result and the leading contribution to the partition function 
at large $N$ are given by:\footnote{The expressions for $a_1$ and $a_2$ in Eq. (\ref{6b}) are given in the thesis and published by Bitter, see Ref. \cite{Bitter37}, p.149.}
\be \label{6b}
Z(N) = a_1w_1^{N-1} [1 + a_2/a_1 (w_2/w_1)^{N-1}] \simeq a_1w_1^{N-1}\, . 
\ee
In the thermodynamic limit this relates the free energy  per elementary magnet 
with $w_1$ via:
\be     \label{7a}
{\cal F} = -kT \lim_{N\to \infty} \ln Z(N) /N = -kT  \ln w_1 \, . 
\ee
From this Ising calculated the magnetization of the chain per particle  as function of temperature  
\be   \label{7b}
{\cal M} = - \partial {\cal F}/\partial h|_{T} = 
\mu \partial \ln w_1/\partial \alpha|_{T}\, .
\ee
Substituting (\ref{6}) into (\ref{7b}) one arrives at the expression
for magnetization that Ising obtained in his thesis
\be\label{9}
{{\cal M}} = \mu \frac{\sinh \alpha}{\sqrt{\sinh^2 \alpha+e^{-2\beta}}} .
\ee
and that brings about absence of spontaneous magnetization at any non-zero
temperature: ${\cal M} (\alpha=0)=0$. 
A more complete analysis of other thermodynamic quantities using his method can be found in the textbook on ferromagnetism by Francis Bitter \cite{Bitter37}, who had access also to the thesis.

\subsection{Reformulation of the Ising problem with the Hamiltonian and transfer matrix for the two-state model}\label{II.3} 

For the sake of completeness, let us now briefly summarize the main
steps of the transfer-matrix solution of the 1D Ising model \cite{Stanley87}.
Here, the starting point is the Hamiltonian for the elementary magnets, which meanwhile have been identified as the electron spins \cite{Pauli30}:
\be
{\cal H} = -J \sum_{j=1}^{N}  S_j S_{j+1} - h \sum_{j=1}^{N}  S_j,
\label{10}
\ee
where $S_j=\pm 1$ are the spin variables, $h$ is an external magnetic field,
$N$ is number of chain sites. This Hamiltonian allows to write down the energy of a configuration as function of the individual states of the electrons (the former elementary magnets, note that the scale of the magnetic moment has been set to one, $\mu=1$) interacting via exchange interaction \cite{Heisenberg28}. The partition function reads:
\begin{equation}
Z(N) = \sum_{\{\rm states \}} e^{- {\cal H}/(kT)} = \sum_{\{\rm states \}}  e^{E_J \sum_{j=1}^{N} S_j
S_{j+1} + \alpha \sum_{j=1}^{N} S_j}
\label{11} 
\end{equation}
where $E_J= J/(kT), \quad \alpha= h/(kT)$, and the sum
\be \label{12}
\sum_{\{\rm states \}} (...) = \prod_{i=1}^N\sum_{S_i=\pm 1} (...)
\ee
means summation 
over the spin states on all
sites. The expression for the partition function can be written in the
form of terms each depending only on two neighbouring spins imposing periodic boundary conditions 
$S_{N+1}= S_1$:
\begin{equation}
Z(N) =  \sum_{\{ \rm states \}}  V(S_1,S_2) V(S_2,S_3) \dots
V(S_{N-1},S_N) V(S_N,S_1),
\label{13}
\end{equation}
with
\begin{equation}
V(S,S^{\prime})=e^{\frac{\alpha}{2}S + E_JS S^{\prime} +
\frac{\alpha}{2}S^{\prime}}.
\label{14}
\end{equation}
As long as $S = \pm 1$, $V(S,S^{\prime})$ takes on four values,
$V(+1,+1)$, $V(+1,-1)$, $V(-1,+1)$, $V(-1,-1)$
that can be conveniently represented as the elements of the
so-called transfer matrix:
\be
{\bf V} = \left ( \begin{array}{ll}
V(+1,+1) & \quad V(+1,-1)
\\
V(-1,+1) & \quad V(-1,-1)
\end{array}
\right )  \quad = \quad
\left ( \begin{array}{ll}
e^{E_J+\alpha} & \quad e^{-E_J}
\\
e^{-E_J} & \quad e^{E_J-\alpha}
\end{array}
\right ).
\label{15}
\ee
Note that for the chain with nearest-neighbour interaction the dimension of the matrix {\bf V} is defined by the number of states taken by the spin $S$.  For the two-state spin variable the matrix is two by two and its elements are the Boltzmann weights of all possible configurations of the two neighbouring spins.
Now, the successive summation  over $S_2, S_3, \dots , S_N$ 
in Eq. (\ref{13}) can be
regarded as a successive matrix multiplication. As a result we
get:
\begin{equation}
Z(N)= {\rm Sp} ({\bf V})^N=({\rm Sp} {\bf V})^N \, .
\label{16}
\end{equation}
The trace (\ref{16}) is equal to the sum of matrix
eigenvalues, which for the matrix ${\bf V}^N$ are equal to
$\lambda_{1,\,2}^N$, with  $\lambda_{1,\,2}$ being the eigenvalues of the
matrix ${\bf V}$ (\ref{15}). 
The eigenvalues are the solutions of the characteristic (secular) equation of the matrix ${\bf V}$
\be
\lambda^2 - 2\lambda e^{E_J}\cosh\alpha + 2\sinh2E_J = 0\, .
\ee
The values readily follow:
\begin{equation}
\lambda_{1, \, 2} = e^{E_J}[ \cosh \alpha \pm \sqrt{ \sinh^2 \alpha + e^{-4E_J}}\,] \, .
\label{17}
\end{equation}
The transfer matrix's largest eigenvalue $\lambda_1$ defines thermodynamics
of the Ising chain. The corresponding functions are expressed in terms
of $\lambda_1$ similar as they were expressed in terms of $w_1$ in the former subsection,
cf. Eqs. (\ref{6b}),  (\ref{7a}),  (\ref{7b}). In particular,
one obtains for the magnetization:
\be\label{18}
{{\cal M}} =\mu
\partial \ln \lambda_1/\partial \alpha|_{T} = \mu
 \frac{\sinh \alpha}{
	\sqrt{
		\sinh^2 \alpha+e^{-4E_J}
	}
} .
\ee
Note the difference in the second terms  
in denominators of Eqs. (\ref{9}) and (\ref{18}): $2E_\beta$ vs $4E_J$. This is due to the fact that
the energy gap for parallel and antiparallel orientation of two neighbouring
 magnetic moments in the original Ising model (subsection
\ref{II.2}) equals $e$, whereas it is equal to $2J$ for the Hamiltonian (\ref{10}).

In order to see the agreement with Ising's result one has to observe that the corresponding transfer matrix has to be modified
\be
{\bf V} \quad = \quad
\left ( \begin{array}{ll}
e^{\alpha} & \quad e^{-2E_J}
\\
e^{-2E_J} & \quad e^{-\alpha}
\end{array}
\right ) \quad = \quad
\left ( \begin{array}{ll}
A_1 & \quad B
\\
B& \quad A_2
\end{array}
\right )
\label{15a}
\ee
since in Ising's energy scale the energy value zero was chosen if the neighbouring spins are parallel. The notations of Eq. (\ref{2}) have been used in the second equality to  emphasize the appearance of the Boltzmann weights. The characteristic polynomial therefore is
\be
\lambda^2- 2\lambda \cosh\alpha + 1-e^{4E_J}=0 \, 
\ee
or
\be
\lambda^2-(A_1+A_2) \lambda + A_1A_2(1-B^2)=0 \, .
\ee
These polynomials are to be compared with  Eqs. (\ref{5b}) or (\ref{5a}) which shows that
 $\lambda$ can be identified with the inverse roots $w$: $\lambda_{1,\, 2}=w_{1,\, 2}$.

 The exact result for the finite chain reads
 \be
 Z(N)=\lambda_1^N(1+(\lambda_2/\lambda_1)^N)
 \ee
 the difference to Eq. \ref{6b} is due to the different boundary conditions and diminishes in the thermodynamic limit.  It should also be remarked that Ising could not calculate correlation functions since the Hamiltonian and the very nature of interacting elements (spin of the electrons) were found after he finished his thesis (for more on the difference caused by different boundary conditions see Chapter III in Ref. \cite{McCoyWu2001}).

\section{Ising's solution of the three-state model}\label{III} 

Ising thought that the disappointing result of his search of a ferromagnetic phase in the chain was due to the `too great idealization'. But a calculation of a spatial model within dimension two or three seemed not to be feasible. Therefore, in the second part of his thesis 
\cite{Ising24}, named `Complicated cases', Komplizierte F\"alle ({\it Germ.}), see Fig. \ref{fig1},
he tried to improve the chain model. In a first step he enlarged the number of states possible for the elementary magnets by allowing the so-called `transverse states' considering them to be perpendicular to the direction of the up and down states and keeping the nearest neighbour interaction. Moreover, for symmetry reasons he allowed $r$ different directions of this transverse states.\footnote{He thinks of the sixfold axis of the pyrrhotite or of the fourfold symmetry axis in magnetite.}  Since the value of $r$ appears in the following calculation only as a trivial parameter not changing the way of treating the three-state model, from now on its value is taken as one, $r=1$.

The way Ising used to calculate the partition function of the three-state model follows closely the steps made for the two-state model. 
This model now has three energy places describing the energy between neighbouring elementary magnets pointing up and down, $e_{12}$, up and transverse, $e_{13}$ and down and transverse, $e_{23}$. Due to the obvious symmetry in interactions the relation $e_{ij}=e_{ji}$ holds  (see also the Appendix). In an external magnetic field $h$ corresponding to the up and down direction, the up and down states attain the energy $\pm \mu h$. In order to prevent turning of the transverse states in the external field, Ising introduced for them an additional, field-independent external energy $e_e$. 

The partition function is expressed by the number of configurations $\cal{N}$ and Boltzmann weights
analogous to Eqs. (\ref{2}) and (\ref{2a})
\bea \nonumber
 Z(N) =\sum_{\nu_1+\nu_3+\nu_3=N}\sum_{{\sigma_{12}=0}}^{N-1} \sum_{{\sigma_{13}=0}}^{N-1} \sum_{{\sigma_{23}=0}}^{N-1}&& \mathcal{N}(\nu_1,\nu_2,\nu_3,\sigma_{12},\sigma_{13},\sigma_{23}) \times\\ 
 &&A_1^{\nu_1}A_2^{\nu_2}A_3^{\nu_3}B_{12}^{\sigma_{12}}B_{13}^{\sigma_{13}}B_{23}^{\sigma_{23}}     \label{partition2}
\eea
where  $\nu_1$, $\nu_2$, and $\nu_3$ are the numbers of up, down, and transverse states correspondingly with the obvious condition $\nu_1+\nu_2+\nu_3=N$ and $\sigma_{12}$, $\sigma_{13}$, and $\sigma_{23}$ are the numbers of respective energy places. The Boltzmann weights read
\be \label{bw}
A_1=  e^{\alpha}, \, A_2= e^{-\alpha},\, A_3=e^{-e_e/(kT)}, \,  B_{ij} = e^{-e_{ij}/(kT)} \, .
\ee
As before, when the neighbours are in the same states,  Ising sets the corresponding energy to zero  (the Boltzmann weights are then equal one).

Following similar steps as described in section \ref{II} and introducing the auxiliary function to find the partition function, Eq. (\ref{partition2}), one observes that again the denominator of the auxiliary function is a polynomial, now of the third order. The zeros of the polynomial are defined by the equation
\begin{eqnarray} \label{cp} \nonumber
&&1-x[A_1+A_2+A_3] + \\  
&&x ^2\left[A_1A_2(1-B_{12}^2)+ A_2A_3(1-B_{23}^2)+A_3A_1(1-B_{31}^2)\right]-\\ \nonumber
&&x^3A_1A_2A_3[1-(B_{12}^2+B_{23}^2+B_{31}^2)+2B_{12}B_{23}B_{31}]=0 \, ,
\end{eqnarray}
whereas the maximal solution of this equation defines the free energy in the
thermodynamic limit $N\to \infty$.

\section{Transfer matrix  formulation for Ising's three-state model}\label{IV} 

Since the energy of a configuration depends only on the nearest-neighbour states, the transfer matrix can be written down in the same way as before in the two-state model analogous to Eqs. (\ref{15}) and (\ref{15a}). Now instead of a $2\times2$  matrix one arrives at a $3\times3$ matrix with the elements that depend on the nearest-neighbour states:
\be \nonumber
{\bf V} = \left ( \begin{array}{ccc}
V(up,\, up) &  V(up, \, down)&  V(up, \, transverse)
\\
V(down,\, up) &  V(down, \, down)&  V(down, \, ttransverse)
\\
V(transverse,\, up) &  V(transverse, \, down)& V(transverse, \, transverse)
\end{array}
\right ) \, .
\ee
Inserting the corresponding Boltzmann weights, the transfer matrix takes on the
following form:
\be  
{\bf V} =  \left ( \begin{array}{ccc}
A_1 &  (A_1A_2)^{1/2}B_{12} &  (A_1A_3)^{1/2}B_{31}\\
(A_1A_2)^{1/2}B_{12} & A_2 & (A_2A_3)^{1/2}B_{23}\\
 (A_1A_3)^{1/2}B_{31} & (A_2A_3)^{1/2}& A_3  
\end{array}
\right )  
\ee 
The characteristic equation for the eigenvalues of the matrix reads:
\begin{eqnarray} \nonumber
&&\lambda^3-\lambda^2[A_1+A_2+A_3] + \\ \label{cpl} 
&&\lambda\left[A_1A_2(1-B_{12}^2)+ A_2A_3(1-B_{23}^2)+A_3A_1(1-B_{31}^2)\right]-\\ \nonumber
&&A_1A_2A_3[1-(B_{12}^2+B_{23}^2+B_{31}^2)+2B_{12}B_{23}B_{31}]=0 \, .
\end{eqnarray}
As in the case of the two-state model, the equation for the transfer-matrix eigenvalues 
and for the inverse roots of the three-state model coincide with each other,
cf. Eqs. (\ref{cpl}) and (\ref{cp}).
Note that only the energies present in the Boltzmann weights are to be defined to get the corresponding eigenvalues for calculating the partition function. In his thesis \cite{Ising25}, Ising generalized this method also for the cases of two chains and for a chain with next nearest neighbours,
cf. paragraphs 7 and 8 in the thesis table of contents displayed in  Fig. \ref{fig1}:
§ 7. The double chain with simultaneous action of adjacent elements of the same
and different chains --
Die Doppelkette bei gleichzeitiger Wirkung benachbarter Elemente derselben
und verschiedener Ketten ({\it Germ.}), 
§ 8. The linear chain at interaction between first and
second adjacent elements -- Die lineare Kette bei Wechselwirkung 
zwischen erst- und zweit-benachbarten Elementen ({\it Germ.}). 
The problem for Ising was that already for the three-state case the characteristic equation is a polynomial of higher than  second order and in order to calculate the partition function in the thermodynamic limit one has to know its largest solution.  In the cases mentioned, this can be done only by solving the equation numerically.

\section{Conclusions and further developments}\label{V}

In his original approach, Ernst Ising used combinatorial methods to calculate 
weights of different elementary magnets configurations and their contributions to the 
partition function. To this end he used an Ansatz introducing an auxiliary function, Eq. (\ref{4}),
 leading to a polynomial whose roots allowed to calculate the system thermodynamics. In particular,
the largest root of the polynomial gives asymptotically  exact expression for the partition function.
Comparing Ising's calculation of the partition function with the analysis of the same model 
by the transfer-matrix technique, a method discovered much later, shows that both methods 
lead to the same characteristic polynomials. In particular, we show in this paper that
the `Ising's roots' arising within the combinatorial treatment  can be identified as the eigenvalues of the 
transfer matrix. 

In 1974, explaining the beginning of the Potts model, Cyril Domb wrote \cite{Domb74}:  `In 1951 when the present author was at Oxford he pointed out to his research student
R B Potts that the transformation discovered by Kramers and Wannier (1941) for the
two-dimensional Ising model could be generalized to a planar vector model having
three symmetric orientations at angles of $0$, $2\pi/3$, $4\pi/3$ with the axis. Hence the Curie
temperature could be located for this model. He suggested that it might be possible to
extend the result to a planar vector model with $q$ symmetric orientations.
After a detailed investigation Potts (1952) came to the conclusion that the transformation
did not generalize to a planar vector model with $q$ orientations, but instead
to a $q$-state model in which there are two different energies of interaction which correspond
to nearest neighbours being in the same state or different states.'

Perhaps, when it comes to a model with a multi-component discrete order parameter, 
the Potts model, a brief history of which is sketched above, comes to mind first.
However, as we emphasize in this paper, an attempt to increase the order parameter component
number
 was already contained in Ising's thesis, carried out almost 30 years earlier 
\cite{Ising24}. These results were not included in his 1925 paper \cite{Ising25} 
and are therefore less well known. Ernst Ising thesis not only analyses 
what is called today the Ising model but 
also it contains the description of the three-state model, which can 
be considered as a  forerunner to 
the models with many-component order parameter, the Potts model of 1952 \cite{Potts52} 
being one of them.  

Meanwhile different variants of $q$-state models have been investigated even for cooperative phenomena outside magnetism.\footnote{See exercises in Ref. \cite{Yeomans92}, p.75.} E.g. the classical spin-1 Ising model is more suitable to describe phase
transitions and critical phenomena occurring in physical systems characterized by three states and such a model has been suggested in 1966 by Blume and Capel for magnetic phase transitions \cite{Blume1966,Capel66}. Later
in 1971 it was extended by Blume, Emery, and Griffiths and used to describe the phase separation in He3–He4 mixtures \cite{Blume1971}, see also \cite{Moueddene24} for recent discussion. Three-state models are  popular in description of biological, economic, social phenomena. Depending on the phenomenon under consideration, obvious interpretations mean  three-state oppositions like buy-sell-hold, susceptible-infected-recovered, left wing-centre-right wing, etc.

At the time of Ising's thesis the importance of the spacial dimensionality of a physical system for the existence of a phase transition was not known. Modern understanding of such phenomenon assumes spatial dimensionality, along with symmetry, order parameter component number, and the interaction
range as key factors determining the class of universality of the system under consideration. 
Renormalization group theory proved the existence of a lower critical dimension for models within a certain universality class below which no ordered phase is possible at non-zero temperature. Although,
the  transition at $T=0$ can be regarded as a critical point \cite{Fisher1983}.
Accordingly, for the problem considered by Ising, such a symmetry group is a discrete group $L_2$, and the corresponding lower critical dimension is $d=1$. Therefore. no spontaneous magnetization can be observed
for this model at non-zero temperature at $d=1$ -- the fact that was confirmed by Ising's exact solution. An absence of 
ordering at non-zero temperature for $d=1$  classical short-range interacting models is 
attributed to an entropy excess relative to interaction energy. 
Notoriously,  for the $d=1$ Ising model,
the entropy-energy balance can be achieved by considering the so-called invisible states \cite{Krasnytska23} that, under certain conditions \cite{Sarkanych17,Sarkanych18} can lead to entropy decrease and thus to promote the spontaneous ordering.

This work has been done in the frames of a larger project that aims at bilingual commented publication of Ernst Ising doctoral thesis \cite{we_four}. We deeply acknowledge our long-standing and enjoyable collaboration with Ber\-trand Berche and Ralph Kenna. We devote this paper to the memory  of Ralph Kenna, our dear friend who recently left us, not even reaching his sixtieth birthday.

Funding: Y.H. acknowledges the support of the JESH mobility programme of the Austrian Academy
of Sciences and hospitality of the Complexity Science Hub Vienna when finalizing this paper.

\begin{appendix}
\section{Schottky's idea and transverse states}

At the time when Ising wrote his thesis, not only the model Hamiltonian and the transfer-matrix method  were unknown. 
Also the very mechanism of interaction leading to the appearance of the low-temperature ferromagnetic phase was a mystery. 
The dipole interaction, which was known at that time, is too weak to explain the observed values of the Curie temperature
({$T_c\sim 1000$ K} for Co, Fe, Ni), whereas the discovery of the exchange interaction, as well as the very concept of spin, were still 
to come, see Ref.\cite{Folk22b} for a more detailed discussion.
In the introduction to the thesis, Ising mentions Schottky's reasoning \cite{Schottky22} about possible physical basis for this 
interaction. However, reference to Schottky is absent in Ising's paper \cite{Ising25} written based on the materials of the thesis.
In this appendix, we explain in more details Schottky's views as expressed in his paper Ref. \cite{Schottky22} and 
show how they may be related to the three-state model considered by Ising.

\begin{figure}[h] 
\begin{center}\includegraphics[height=5.0cm]{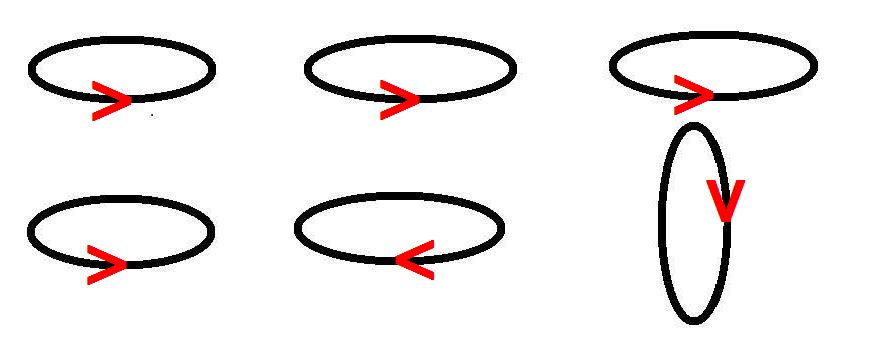}\end{center}
\centerline{\bf (a) \hspace{7em} (b) \hspace{7em} (c)}
\caption{Different configurations of two neighboring rotating electrons (red arrows show direction of rotation). Each of the electrons can be in one of the three states: rotating in a parallel plane clockwise (state `+' or `up'),
counterclockwise (state `-' or `down'), or rotating in a perpendicular plane (state `0' or `transverse').
 The direction of the chain is downwards and coincides with the direction of the external field if present.
Corresponding Boltzmann weights are given in Eq. (\ref{bw}). The configuration (b) has the weight $B_{12}$ and (c) $B_{23}$ (or $B_{13}$ (according to Schottky's calculation).  \label{fig3}}
\end{figure}

 Schottky has formulated his ideas in terms
of the old Bohr-Sommerfeld picture of quantum mechanics. He considers circling electrons on neighbouring atoms as shown
in Fig. \ref{fig3}, panels {\bf (a)} and {\bf (b)}. Because of the rotation, the electrons produce elementary
magnetic moments, pointing along the axis, perpendicular to the rotation plane. Therefore,  the direction of the chain 
of elementary magnets is downwards in the Fig. \ref{fig3}. 
When the electrons are circling in the same directions, panel {\bf (a)}, the induced magnetic moments are parallel.
The opposite directions, panel {\bf (b)}, corresponds to the antiparallel orientation of
the induced magnetic moments.
The mean energy of the (electrostatic) interaction $E$ depends on the distance $d$ 
between the electrons. The latter, in turn, depends both on the phase $\phi$ and the direction of rotation. 
The key concept of 
Schottky's theory is a {\it synchronism} in the motion of the circling electrons: 
the energy $E$ of the two electrons should be as small as possible during the circling around the nucleus.
It is the phase between the two circling electrons which is adjusted
to minimize the energy of the Coulomb interaction. 
The preferred, parallel or antiparallel, configuration of magnetic moments is defined by the 
difference between their electrostatic energies. The difference in the energy due to the magnetic dipole
interaction might be neglected, being much smaller than the electrostatic energy difference.

For illustration purposes, similar as in Ref. \cite{Folk22b}, let us calculate the distance 
between  two electrons rotating with frequency $\omega$ along two circles of radius $r$ placed 
above each other at distance $a$ (see Fig. \ref{fig3}{\bf a}).  The phase between the two 
rotations is fixed to $\phi$. The coordinates of two electrons read
 \begin{align}
x_1&= r \cos\omega t & x_2&= r \cos(\omega t+\phi)\\
y_1&=r \sin\omega t  & y _2&=r \sin(\omega t+\phi)\\
z_1&=0 & z_2&=a \, .
\end{align}
Introducing the notation $\omega t=\tau$, $R=r/a$  we get for the distance if both electrons rotate in the same direction:
\begin{equation}
d_{1}(R,\tau,\phi)=\sqrt{1+4R^2\sin^2(\phi/2)}\, .
\end{equation}
Now changing the direction of circulation of one electron [$\tau\to-\tau$ ], cf. Fig. \ref{fig3}{\bf b}, 
one gets for the distance
\begin{equation}
d_{2}(R,\tau,\phi)=\sqrt{1+4R^2\sin^2(\tau+\phi/2)} \, .
\end{equation}

\begin{figure}\centering
\includegraphics[width=0.5\textwidth]{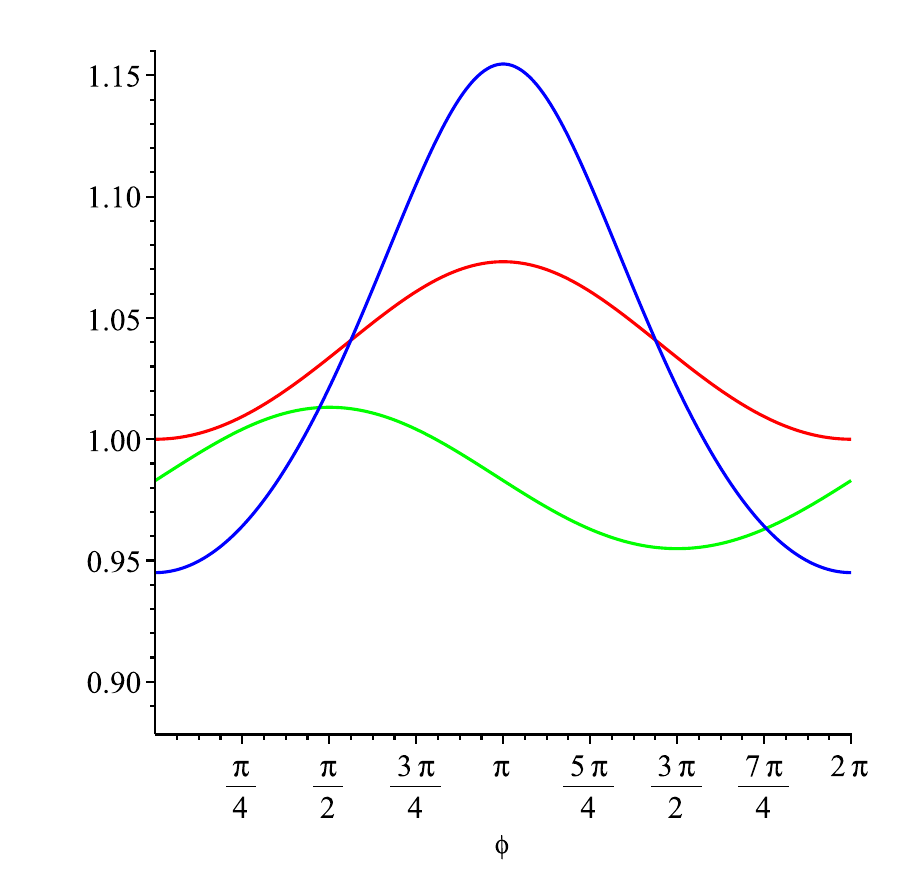}\includegraphics[width=0.5\textwidth]{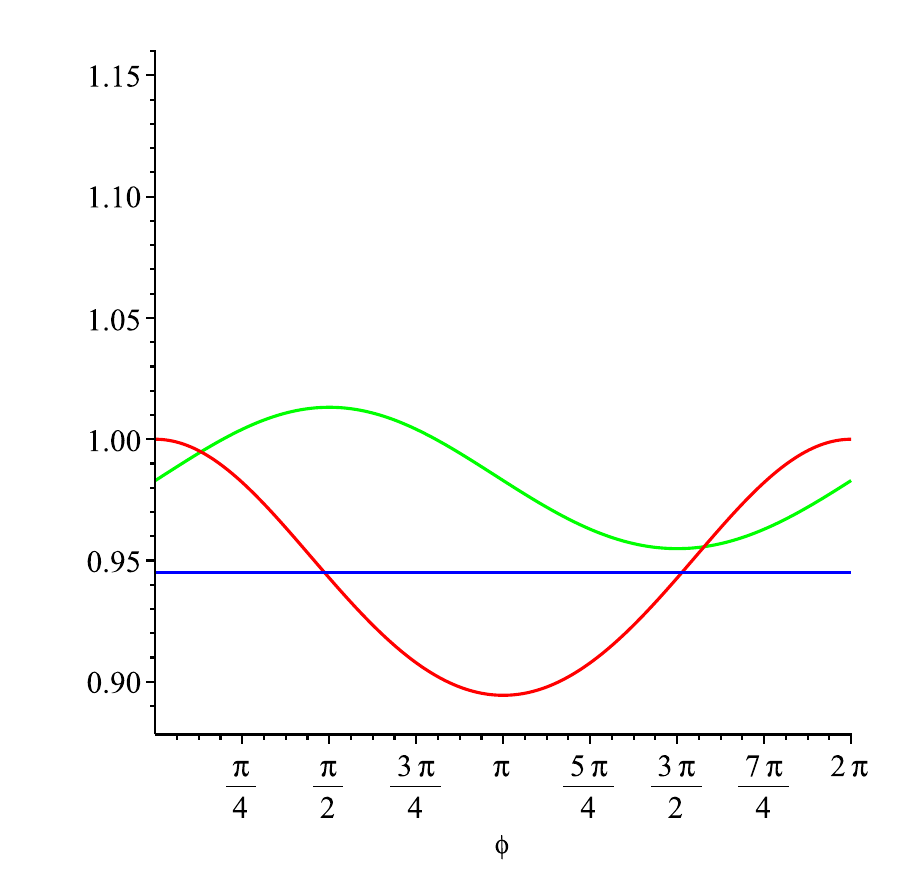}
\centerline{\bf (a) \hspace{12em} (b)}
\caption{Mean interaction energies of two rotating electrons of 
different configurations as functions of phases. {\bf (a)}: electrons are placed beside each other,
{\bf (b)}: electrons are placed above each other. Colours show the states of the electron pairs:
both electrons rotate in parallel planes in the same (red) and in opposite (blue)
directions,
one electron rotates in a parallel and another one in a perpendicular plain (green). 
Note that in the latter case  the mean interaction energy does not depend on the rotation 
direction. \label{fig4}}
\end{figure}

Consider now one electron rotating in the $xy$ plane (as in the former case), 
the other one rotating in the $zy$ plane, as shown in Fig. \ref{fig3}{\bf c}
(direction of rotation does not matter in this case). The coordinates read
 \begin{align}
x_1&= r \cos(\omega t) & x_2&= 0\\
y_1&=r \sin(\omega t) & y _2&=r \cos(\omega t+\phi)\\
z_1&=0 & z_2&=r \sin(\omega t+\phi)+a\, .
\end{align}
We get for the distance
\begin{equation}
d_{3}(R,\tau,\phi)=\sqrt{2R^2+1\pm 2R\sin(\tau)-2R^2\cos(\tau)\sin(\phi\pm \tau)} \, .
\end{equation}

Here, signs $\pm$ correspond to different directions of electron rotation. Note however,
that all four combinations of $\pm$ signs lead to the same values of the mean energy.

The mean Coulomb energy for different configurations of rotating electrons 
can be defined as:
\be
E_i(R,\phi)=\frac{1}{2\pi}\int_0^{2\pi}\frac{d\tau}{d_i(R,\tau,\phi)}\, .
\ee

Repeating similar reasoning one gets mean energies for the cases,
when electrons are circling in the planes beside (and not above) each other.
The resulting mean energy plots  are shown in Fig. \ref{fig4}. 
As one can see from the plots, the analysis of the three-state model within the framework of Schottky's  synchronism concept leads to a similar conclusion as in the case of the two-state model \cite{Schottky22,Folk22b}: the existence of configurations characterized by the minimum average interaction energy. In this way the model had a still weak microscopic (physical?) basis.

\end{appendix}

\end{document}